\documentclass[prl,twocolumn,nofootinbib]{revtex4}
\usepackage{graphicx, epsfig}
\usepackage{color}
\usepackage{mathrsfs}

\usepackage{amsmath}
\usepackage{amssymb}

\def \lleq {\lower0.9ex\hbox{ $\buildrel < \over \sim$} ~}
\def \ggeq {\lower0.9ex\hbox{ $\buildrel > \over \sim$} ~}

\newcommand{\ben}{\begin{eqnarray}}
\newcommand{\een}{\end{eqnarray}}

\def \beq  {\begin{equation}}
\def \eeq  {\end{equation}}
\def \ber  {\begin{eqnarray}}
\def \eer  {\end{eqnarray}}

\newcommand {\ga} {\ {\raise-.5ex\hbox{$\buildrel>\over\sim$}}\ }
\newcommand {\la} {\ {\raise-.5ex\hbox{$\buildrel<\over\sim$}}\ }

\begin{document}
\newcommand{\newc}{\newcommand}

\newc{\be}{\begin{equation}}
\newc{\ee}{\end{equation}}
\newc{\ba}{\begin{eqnarray}}
\newc{\ea}{\end{eqnarray}}
\newc{\bea}{\begin{eqnarray*}}
\newc{\eea}{\end{eqnarray*}}
\newc{\D}{\partial}
\newc{\ie}{{\it i.e.} }
\newc{\eg}{{\it e.g.} }
\newc{\etc}{{\it etc.} }
\newc{\etal}{{\it et al.}}
\newc{\lcdm }{$\Lambda$CDM }
\newcommand{\nn}{\nonumber}
\newc{\ra}{\rightarrow}
\newc{\lra}{\leftrightarrow}
\newc{\lsim}{\buildrel{<}\over{\sim}}
\newc{\gsim}{\buildrel{>}\over{\sim}}
\title{PPN Parameter $\gamma$ and Solar System Constraints of Massive Brans-Dicke Theories}
\author{L. Perivolaropoulos}
 \affiliation{Department of Physics, University of Ioannina, Greece}
\date{\today}

\begin{abstract}
Previous solar system constraints of the Brans-Dicke (BD) parameter $\omega$ have either ignored the effects of the scalar field potential (mass terms) or assumed a highly massive scalar field. Here, we interpolate between the above two assumptions and derive the solar system constraints on the BD parameter $\omega$ for {\it any} field mass. We show that for $\omega=O(1)$ the solar system constraints relax for a field mass  $m \gsim 20 \times m_{AU}= 20\times 10^{-27}GeV$.
\end{abstract}
\maketitle

Scalar-Tensor (ST) theories \cite{st-action} constitute a fairly generic extension of General Relativity (GR) where the gravitational constant is promoted to a field whose dynamics is determined by the following action \cite{st-action,EspositoFarese:2000ij}
\begin{widetext}
\be
S={1\over 16\pi G} \int d^4x \sqrt{-g}
\Bigl(F(\Phi)~R -
Z(\Phi)~g^{\mu\nu}
\partial_{\mu}\Phi
\partial_{\nu}\Phi
- 2U(\Phi) \Bigr)
+ S_m[\psi_m; g_{\mu\nu}]\ .
\label{action}
\ee
\end{widetext}
where G is the bare gravitational constant, $R$ is the scalar curvature of the metric $g_{\mu\nu}$ and $S_m$ is the action of matter fields. The variation of the dimensionless function $F(\Phi)$ describes the variation of the effective gravitational constant. This variation (spatial or temporal) is severely constrained by solar system experiments \cite{EspositoFarese:2004cc,Will:2001mx,cassini}. The GR limit of ST theories is obtained either by fixing $F(\Phi)=\Phi_0\simeq 1$ ($\Phi_0$ is a constant) or by freezing the dynamics of $\Phi$ using the function $Z(\Phi)$ or the potential $U(\Phi)$. For example a large and steep $Z(\Phi)$ makes it very costly energetically for $\Phi$ to develop a kinetic term while a steep $U(\Phi)$ (massive $\Phi$) can make it very costly energetically for $\Phi$ to develop potential energy. In both cases we have an effective {\it freezing} of the dynamics which reduces the ST theory to GR.

ST theories have attracted significant attention recently as a potentially physical mechanism\cite{EspositoFarese:2000ij,Boisseau:2000pr,Torres:2002pe} for generating the observed accelerating expansion of the universe (see Ref. \cite{Copeland:2006wr,Perivolaropoulos:2006ce} and references therein). A significant advantage of this mechanism is that it can naturally generate an accelerating expansion rate corresponding to an effective equation of state parameter $w_{eff}$ that crosses the phantom divide line $w=-1$ \cite{Boisseau:2000pr,Perivolaropoulos:2005yv,Nesseris:2006er}. Such a crossing is consistent with cosmological observations and is difficult to obtain in the context of GR \cite{Nesseris:2006hp}. In addition ST theories naturally emerge in the context of string theories\cite{dilaton} and in Kaluza-Klein\cite{kal-kl} theories with compact extra dimensions\cite{Perivolaropoulos:2002pn}.

A special case of ST theories is the Brans-Dicke (BD) theory\cite{bdtheor} where \ba F(\Phi)&=&\Phi \\ Z(\Phi)&=&\frac{\omega}{\Phi} \label{bddef}\ea For a massive BD theory we also assume a potential of the form \be U(\Phi)=\frac{1}{2} m^2 (\Phi-\Phi_0)^2 \label{bdpot} \ee Clearly, the spatial dynamics of $\Phi$ can freeze for $\omega \gg 1$ or for $m \gg r^{-1}$ where $r$ is the scale of the experiment or observation testing the dynamics of $\Phi$. For solar system scale observations, the relevant scale is the Astronomical Unit ($AU\simeq 10^8 km$) corresponding to a mass scale $m_{AU} \simeq 10^{-27}GeV$. Even though this scale is small for particle physics considerations, it is still much larger than the Hubble mass scale $m_{H_0}\simeq 10^{-42} GeV$ required for non-trivial cosmological evolution of $\Phi$\cite{Albrecht:2001xt,Torres:2002pe}.

Current solar system constraints\cite{ppn,Will:2001mx} of the BD parameter $\omega$ have been obtained under one of the following assumptions:
\begin{itemize}
\item {\bf Negligible mass of the field $\Phi$ ($m \ll m_{AU}$):} In this case the relation between the observable Post-Newtonian parameter $\gamma$ (measuring how much space curvature is produced by a unit rest mass)\cite{ppn} and $\omega$ is of the form \cite{Will:2001mx,Xu:2007dc,Klimek:2009zz} \be \gamma(\omega)=\frac{1+\omega}{2+\omega} \label{gambd1} \ee This relation combined with the solar system constraints of the Cassini mission \cite{cassini} \be \gamma_{obs} - 1 = (2.1 \pm 2.3)\times 10^{-5} \label{gamobs} \ee which constrain $\gamma$ close to its GR value $\gamma=1$, leads to the constraint on $\omega$ \be \omega > 4\times 10^4 \label{omcons} \ee at the $2\sigma$ confidence level. Equation (\ref{gambd1}) however should not be used in the case of massive BD theories as was attempted recently in Ref. \cite{Capone:2009xk}.
    \item  {\bf Very massive scalar field $\Phi$ ($m\gg m_{AU}$):} In this case the spatial dynamics of $\Phi$ is frozen on solar system scales by the potential term and all values of $\omega$ are observationally acceptable even though rapid oscillations of the field can lead to interesting non-trivial effects\cite{Wagoner:1970vr,Steinhardt:1994vs,Perivolaropoulos:2003we}. \end{itemize}
In this study we fill the gap between the above two assumptions and derive the form of the predicted {\it effective} parameter $\gamma$ for all values of the field mass $m$. In particular, we derive the form of $\gamma(\omega,m,r)$ where $r$ is the scale of the experiment-observation constraining $\gamma$. We then use the current solar system constraints (\ref{gamobs}) to obtain the $(\omega,m)$ parameter regions allowed by observations at the $1\sigma$ and $2\sigma$ confidence level.

The dynamical equations obtained for the field $\Phi$ and the metric $g_{\mu\nu}$ by variation of the action (\ref{action}) in the massive BD case defined by equations (\ref{bddef}), (\ref{bdpot}) are of the form \cite{EspositoFarese:2000ij}
\begin{widetext}
\be
\Phi \left(R_{\mu\nu}-{1\over2}g_{\mu\nu}R\right)
= 8\pi G T_{\mu\nu}
+ \frac{\omega}{\Phi} \left(\partial_\mu\Phi\partial_\nu\Phi
- {1\over 2}g_{\mu\nu}
(\partial_\alpha\Phi)^2\right)+\nabla_\mu\partial_\nu F(\Phi) - g_{\mu\nu}\Box \Phi
- g_{\mu\nu} \frac{1}{2} m^2 (\Phi-\Phi_0)^2 \label{metdyneq} \ee
\be
(2\omega +3)\Box\Phi =
8\pi G \,T + 2m^2 \left((\Phi-\Phi_0)^2 + (\Phi-\Phi_0)\Phi\right)
\label{phidyneq}
\ee
\end{widetext}
\begin{figure*}[!t]
\centering
\begin{center}
$\begin{array}{@{\hspace{-0.10in}}c@{\hspace{0.0in}}c}
\multicolumn{1}{l}{\mbox{}} & \multicolumn{1}{l}{\mbox{}} \\
[-0.2in] \epsfxsize=3.3in \epsffile{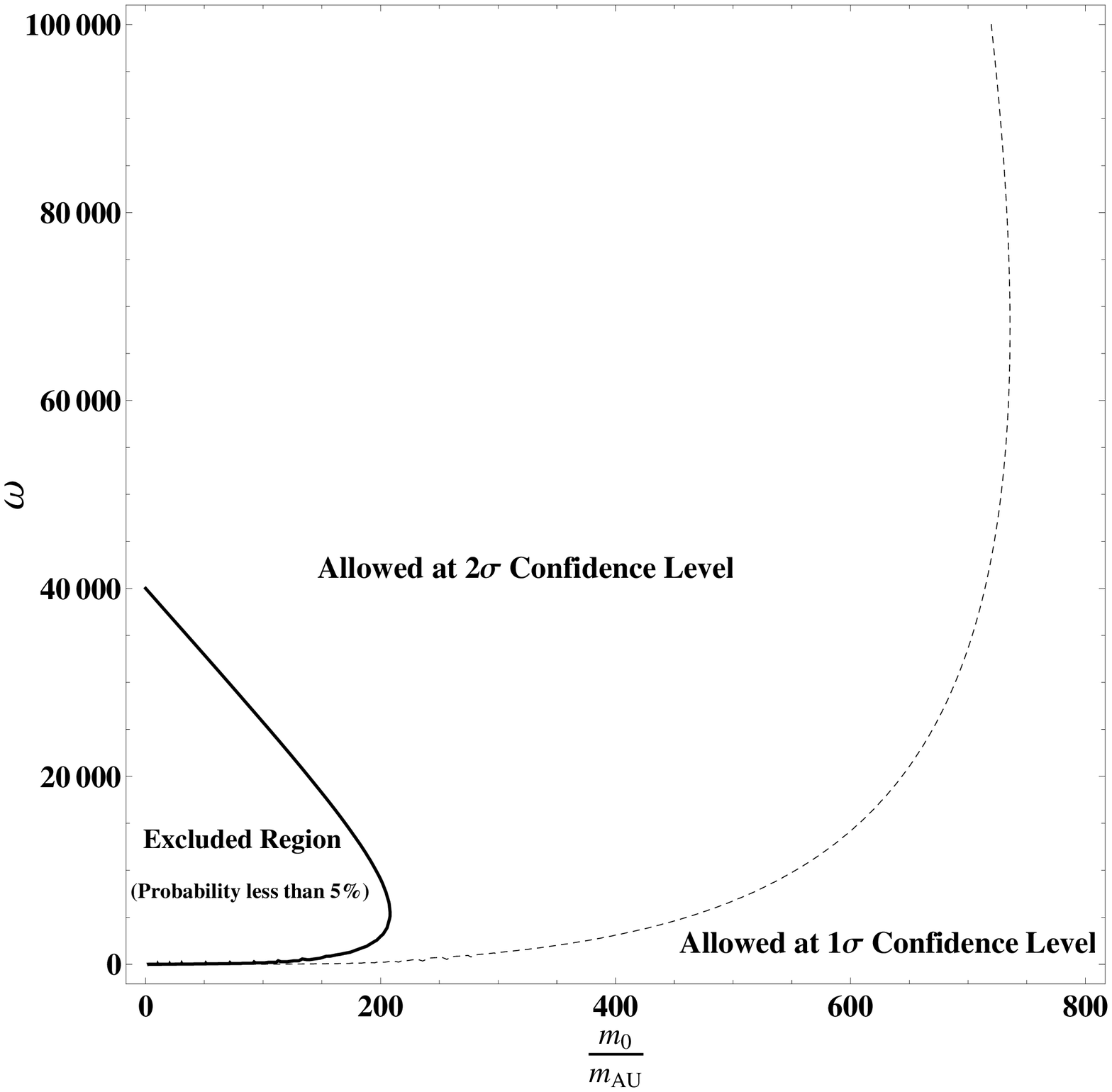} & \epsfxsize=3.3in
\epsffile{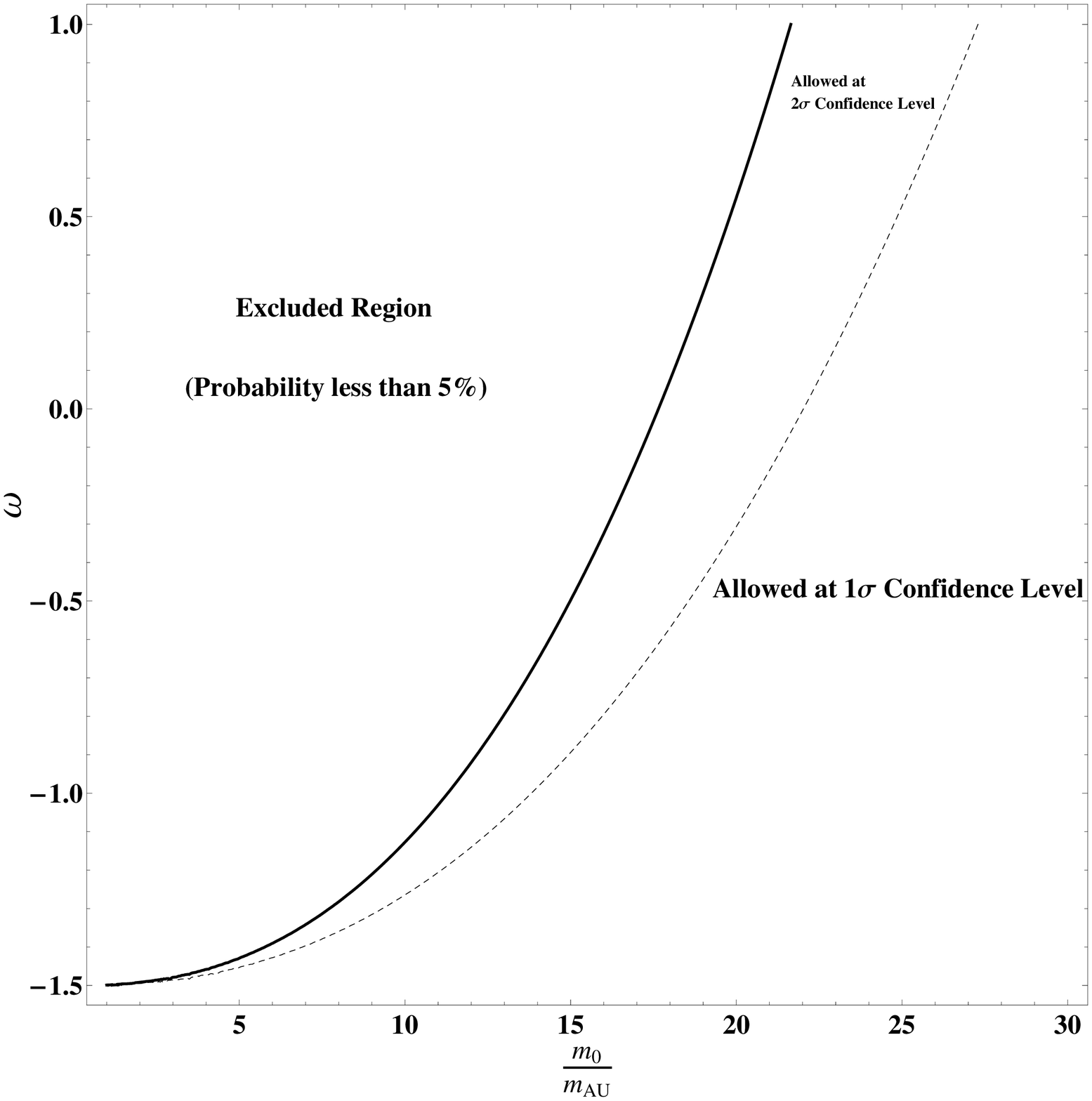} \\
\end{array}$
\end{center}
\vspace{0.0cm} \caption{\small a. The observationally allowed regions for the parameters $\omega$ and $m_0\equiv \sqrt{2\Phi_0} m\simeq m$ at $1\sigma$ 68\% confidence level (above and right of dashed line) and $2\sigma$ 95\% confidence level (above and right of thick line). Notice that for $\frac{m_0}{m_{AU}}\gsim 200$ all values of $\omega$ are observationally allowed at the $2\sigma$ level. b. Same a Fig. 1a focused on a region close to the origin. Notice that for $\omega=O(1)$ solar system constraints relax for $\frac{m_0}{m_{AU}}\gsim 20$ at the $2\sigma$ level (thick line). } \label{fig1}
\end{figure*}
Considering the physical setup of the solar system involving a weak gravitational field we expand around a constant-uniform background field $\Phi_0$ and a Minkowski metric $\eta_{\mu\nu}=diag(-1,1,1,1)$
\footnote{Cosmological considerations would allow a slow evolution of $\Phi_0=\Phi_0(t)$ on cosmological timescales but since these timescales are much larger than the solar system timescales we may ignore that evolution for our physical setup.}
\ba \Phi&=&\Phi_0 +\varphi \label{expf}  \\ g_{\mu\nu}&=&\eta_{\mu\nu}+h_{\mu\nu} \label{expg} \ea
The resulting equations for $\varphi$ and $h_{\mu\nu}$ obtained from (\ref{metdyneq}), (\ref{phidyneq}), (\ref{expg}) and (\ref{expf}) in the gauge $h^\mu_\nu,_\mu-\frac{1}{2}h^\mu_\mu,_\nu=\frac{1}{\Phi_0}\varphi,_\nu$ are
\be
\left(\Box - \frac{2 m^2 \Phi_0}{2\omega+3}\right)\varphi = -8\pi G \frac{\rho - 3p}{2\omega +3} \label{linphieq} \ee
\be -\frac{\Phi_0}{2} \left[\Box (h_{\mu\nu}-\eta_{\mu\nu}\frac{h}{2})\right]= 8\pi G T_{\mu\nu} +\partial_\mu\partial_\nu \varphi-\eta_{\mu\nu}\Box\varphi \label{linmeteq} \ee
where $T_{\mu\nu}=diag(\rho,p,p,p)$ and $h=h^\mu_\mu$. Since we are interested in approximately static solutions corresponding to a gravitating mass such as the Sun or the Earth we ignore time derivatives and set $p\simeq 0$. Thus equations (\ref{linphieq}), (\ref{linmeteq}) become \ba \nabla^2\varphi - \frac{2 m^2 \Phi_0}{2\omega +3} \varphi &=& -8\pi G \frac{\rho}{2\omega +3} \label{linphieq1} \\
\Phi_0 \nabla^2 h_{00} - \nabla^2 \varphi &=& -8\pi G \rho \label{linmeteq1a} \\
\Phi_0 \nabla^2 h_{ij} - \delta_{ij} \nabla^2 \varphi &=& -8\pi G \rho  \delta_{ij} \label{linmeteq1b} \ea
These equations are consistent with corresponding results of Ref. \cite{Wagoner:1970vr,Steinhardt:1994vs,Olmo:2005jd} even though our notation and assumptions are somewhat different. Setting $\rho=M_s \delta(r)$ we obtain the following solution \ba \varphi &=&\frac{2 G M_s}{(2\omega+3)r}e^{-{\bar m}(\omega)r} \label{phisol} \\
h_{00}&=&\frac{2GM_s}{\Phi_0 r}\left(1+\frac{1}{2\omega+3}e^{-{\bar m}(\omega)r}\right)\label{h00sol} \\
h_{ij}&=&\frac{2GM_s}{\Phi_0 r}\delta_{ij}\left(1-\frac{1}{2\omega+3}e^{-{\bar m}(\omega)r}\right)\label{hijsol} \ea where ${\bar m}(\omega)\equiv\sqrt{\frac{2\Phi_0}{2\omega+3}}m$ ($\Phi_0$ is dimensionless). Using now the standard expansion of the metric in terms of the $\gamma$ Post-Newtonian parameter \ba g_{00}&=&-1+2u \label{gamdef1} \\ g_{ij}&=&(1+2\gamma u)\delta_{ij} \label{gamdef2}\ea where $u$ is the Newtonian potential we find (see also \cite{Olmo:2005jd}) \be \gamma(\omega,m,r)=\frac{h_{ij}\vert_{i=j}}{h_{00}}=\frac{1-\frac{e^{-{\bar m}(\omega)r}}{2\omega+3}}{1+\frac{e^{-{\bar m}(\omega)r}}{2\omega+3}} \label{gamfull} \ee
In the special case of $m=0$ we obtain the familiar result of equation (\ref{gambd1}).

The effective mass ${\bar m}(\omega)$ imposes a range ${\bar m}(\omega)^{-1}$ to the gravitational interaction in BD theories. In these theories, the Newtonian potential is \be h_{00}=2u=\frac{2G_{eff} M_s}{r} \label{ntpot1} \ee with \be G_{eff}=\frac{G}{\Phi_0}\left(1+\frac{1}{2\omega+3}e^{-{\bar m}(\omega)r}\right) \label{geff} \ee
The dependence of the effective parameter $\gamma$ on the scale should be interpreted as a dependence on the scale of the experiment-observation imposing a bound on $\gamma$. For example, for solar system constraints we have $r\simeq 1AU\simeq 10^8 km$ which corresponds to the mass scale $m_{AU}\simeq 10^{-27}GeV$. For ${\bar m}(\omega)\gsim m_{AU}$, the value of $\gamma$ predicted by BD theories for solar system scale observations (equation (\ref{gamfull})) is significantly different from the standard expression (\ref{gambd1}). The other major Post-Newtonian parameter $\beta$ (measuring how much `non-linearity' there is in the superposition law of gravity) is not discussed in this study but it is anticipated to remain at its GR value $\beta=1$ as in the case of massless BD theories \cite{Will:2001mx} since the mass term can only improve the consistency with GR.

In order to constrain the allowed $\omega- m$ parameter region we use the recent observational estimates of equation (\ref{gamobs}) obtained by the Cassini spacecraft delay into the radio waves transmission near the solar conjuction \cite{cassini}. Equation (\ref{gamobs}) implies lower bound constraints on the parameter $\gamma$ {\it ie} \ba \gamma(\omega,m,m_{AU}^{-1})&>&1-0.2\times 10^{-5} \label{gamobs1s} \\
\gamma(\omega,m,m_{AU}^{-1})&>&1-2.5\times 10^{-5} \label{gamobs2s} \ea at the $1\sigma$ and $2\sigma$ levels respectively.\footnote{Equations (\ref{gamobs1s}) and (\ref{gamobs2s}) are obtained by subtracting the $1\sigma$ error ($\delta \gamma=2.3\times 10^{-5}$) and the $2\sigma$ error ($2\delta \gamma=4.6\times 10^{-5}$) respectively from the mean value of ${\bar\gamma}_{obs}=1+2.1\times 10^{-5}$ of equation (\ref{gamobs}).} Using equations (\ref{gamfull}) and (\ref{gamobs1s})-(\ref{gamobs2s}) we may find the observationally allowed range of $\omega$ for each value of $m$ (measured in units of $m_{AU}\simeq 10^{-27} GeV$) at the $1\sigma$ and $2\sigma$ confidence levels. This allowed range at $2\sigma$ confidence level is shown in Fig. 1 (regions above and on the right of the thick line). The thick line of Fig. 1 is obtained by equating the expression of $\gamma(\omega,m,r_{AU})=\gamma(\omega,m,m_{AU}^{-1})$ (eq. (\ref{gamfull})) with the $2\sigma$ limit of equation (\ref{gamobs2s}) and plotting the corresponding contour in the $(\omega,m/m_{AU})$ parameter space. The dashed line of Fig. 1 is obtained in a similar way using the $1\sigma$ limit of equation (\ref{gamobs1s}). Clearly, for $\omega=O(1)$ and $\frac{{m}}{m_{AU}}\gsim 20$ the solar system constraints relax and values of $\omega=O(1)$ are allowed by solar system observations at the $2\sigma$ level. For $m=0$ we reobtain the familiar bound $\omega \gsim 40000$ at $2\sigma$ level while for $m\gsim 200 m_{AU}$ all values of $\omega$ are allowed. The plot of Fig. 1 can be used for any experiment-observation constraining the parameter $\gamma$ on a scale $r$ by proper reinterpretation of the units of the $m$ axis.

In conclusion, we have used solar system constraints of the Post-Newtonian parameter $\gamma$ to find the allowed $(\omega,m)$ parameter region of massive BD theories for all values of the scalar field mass $m$ including the mass scale $m_{AU}$ corresponding to the solar system distance scale. This result, fills a gap in the literature where only the cases $m\ll m_{AU}$ and $m\gg m_{AU}$ had been considered. We have found that for $ m\simeq m_{AU}\simeq 10^{27}GeV$, the observationally allowed range of $\omega$ at the $2\sigma$ level is practically identical to the corresponding range corresponding to the $m=0$ range of equation (\ref{omcons}). However, for $m \gsim 200 m_{AU}$ all values of $\omega>-\frac{3}{2}$ are observationally allowed. 

An interesting extension of the present study would be the generalization of the well known expression of $\gamma$ in scalar-tensor theories \be \gamma(F,Z) - 1 = - {(dF(\Phi)/d\Phi)^2\over Z(\Phi) F(\Phi)
+ 2(dF(\Phi)/d\Phi)^2}
\label{gamst} \ee
which like equation (\ref{gambd1}) ignores the possible stabilizing effects of the scalar field potential. Such a generalization would lead to an expression of the effective scale dependent parameter $\gamma$ in terms of $F(\Phi)$, $Z(\Phi)$ and $U(\Phi)$.

\section*{Acknowledgements}
I thank David Polarski, Thomas Sotiriou and Gonzalo Olmo for useful comments. This work was supported by the European Research and
Training Network MRTPN-CT-2006 035863-1 (UniverseNet).

\end{document}